\documentstyle[12pt]{article}

\input epsf
\begin{document}

\newcommand{\xperp}{{\vec{x}_{\perp}}}
\newcommand{\kperp}{{\vec{k}_{\perp}}}
\newcommand{\kdotx}{\kperp \cdot \xperp}
\newcommand{\eq}{\begin{equation}}
\newcommand{\en}{\end{equation}}
\newcommand{\gsi}{\,\raisebox{-0.13cm}{$\stackrel{\textstyle>}
{\textstyle\sim}$}\,}
\newcommand{\lsi}{\,\raisebox{-0.13cm}{$\stackrel{\textstyle<}
{\textstyle\sim}$}\,}

\rightline{RU-95-01}
\baselineskip=18pt
\vskip 0.5in
\begin{center}
{\bf \LARGE CP Violation and the \\ Baryonic Asymmetry of the Universe}
\end{center}
\vspace*{0.5in}
\begin{center}{\large Glennys R. Farrar}\footnote{Research supported
in part by NSF-PHY-91-21039.  Invited talk at Trends in Astroparticle
Physics, Stockholm, Sweden, September 1994, to appear in Nuc. Phys. B
Proc. Suppl. } \\
\vspace{.05in}
{\it Department of Physics and Astronomy \\ Rutgers University,
Piscataway, NJ 08855, USA}
\end{center}
\vspace*{0.8in}

{\bf Abstract:}

The physics of electroweak baryogenesis is described with the aim of
making the essentials clear to non-experts.  Several models for the
source of the necessary CP violation are discussed: CKM phases as in
the minimal standard model, general two higgs doublet models, the
supersymmetric standard model, $Z$ condensates, and the singlet
majoron model.  In a more technical section, a strategy is introduced
for consistently treating quark dynamics in the neighborhood of the
bubble wall, where both local and non-local interactions are
important.  This provides a method for deciding whether gluonic
corrections wash out the elecroweak contribution to the baryonic
asymmetry in the minimal standard model.

\thispagestyle{empty}
\newpage
\addtocounter{page}{-1}
\newpage

\section{General Picture}
\label{general}

A major challenge to particle theory and cosmology is to account for
the tiny but non-zero value of the baryonic asymmetry of the universe
(bau).  The main purpose of this lecture was to give an understandable
yet relatively complete overview of the subject for non-experts, with
special emphasis on the possible sources of CP violation.
Interestingly, it is easier to obtain reliable predictions from
extensions of the minimal standard model (\S \ref{majoron},\ref{2higgs})
than from the minimal standard model (MSM -- \S \ref{msm}).  The major
difficulty is discussed in section \S \ref{qmreflection}.  A method
for overcoming the difficulty is outlined in the final section.

Comparing the results of the theory of nucleosynthesis to observations
on the abundances of primordial light nuclei, the ratio of baryon
number density to entropy is infered to be $\frac{n_B}{s} \sim (2-4)
\times 10^{-11}$\cite{nucleosyn}.  If the universe starts out with no
net baryon number, $B=0$, then baryon number violating processes are
necessary in order for a non-zero baryonic density to be produced.  A
significant baryonic asymmetry can only be produced at a phase
transition, because in thermal equilibrium detailed balance guarantees
that no asymmetry can develop.  It was natural that early work
focussed on a possible GUT (Grand Unified Theory) phase transition
origin for the bau, since by their very nature GUT multiplets contain
both leptons and quarks and thus GUTs naturally have baryon number
violating transitions mediated by the GUT gauge and Higgs particles.

However Kuzmin, Rubakov and Shaposhnikov showed in
1985\cite{krs85,dimsus} that for temperatures above the electroweak
(ew) phase transition temperature, sphaleron processes occur at a
sufficiently high rate to greatly reduce and possibly even wipe out a
baryonic asymmetry produced at the GUT phase transition.  A sphaleron
is a thermal fluctuation in the electroweak gauge fields, which
connects vacuum configurations of different winding number.  It has
similar effects to the instanton, which is a quantum tunneling event
which occurs (at a negligible rate\cite{hooft:prl}) in the
zero-temperature theory.  The energy levels of electroweak doublets
present in the Dirac sea (left-chiral quarks and leptons) shift in
response to changing external gauge fields.  Remarkably, when the
Chern-Simons number (roughly speaking, the number of twists in the
vector potential giving rise to the gauge fields) changes by one unit,
such as occurs in the neighborhood of a sphaleron or instanton, one
complete set of fermions in the Dirac sea is ``promoted'' into being
real particles.  This gives rise\cite{hooft:pr} to an effective
interaction producing or destroying one each of the electroweak gauge
doublets -- i.e., creating or annihilating 9 quarks (one left chiral
quark of each of the three colors and three generations) and three
left-chiral leptons (one from each generation).  In the presence of a
baryon excess, these interactions will proceed in a direction which
reduces the free energy by converting part of the baryon excess to
anti-leptons. Since the sphaleron conserves $B-L$, if GUT phenomena produce
a $B-L$ excess and not simply a $B$ excess, that will not be affected
by the ew sphaleron.  However in SU(5), the most popular GUT, $B-L$ is
conserved so that one must rely on various other effects to circumvent
the sphaleron, or use a more complicated GUT.  While there are a
number of ways that GUT baryogenesis can be made to work in spite of
these sphaleron transitions, interest has shifted to studying the
possibility that phenomena occuring at the ew phase transition
produced the observed bau and that will be the main subject of this
talk.

Production of a non-zero bau requires CP violation as well as baryon
number violation, since if CP is a good symmetry, processes in which
particles are replaced by antiparticles of the same chirality will
occur with equal rates\footnote{Chirality is the same as helicity for
a massless particle, and the opposite of the helicity for a massless
antiparticle.  The CP conjugate of a left chiral particle is a left
chiral antiparticle, etc.}.  It is of course well known that nature
does not fully respect CP symmetry, since the $K^0_L$ has been
observed to decay into both CP even and CP odd final states.  At the
very least, this requires the physical $K^0_L$ to be a superposition
of even and odd CP.  In the minimal standard model (MSM) the CP
violation observed in the kaon system is explained as arising from the
existence of an explicitly CP-violating phase in the
Cabibbo-Kobayashi-Maskawa (CKM) matrix -- the matrix describing mixing
between gauge and mass eigenstates of the quarks.  If the only source
of CP violation is the CKM phase, as is the only possibility in the
minimal standard model, then the magnitude of all other occurences of
CP violation can be predicted because the parameters of the CKM matrix
are fully determined, albeit with limited precision at the present
time\footnote{A more detailed discussion of CKM CP violation will
follow in section \ref{msm}.  For a review of experimental information
on the CKM parameters, see \cite{gn}}.

While CKM CP violation is very popular theoretically, since it appears
naturally in the minimal standard model, it is not excluded that even
the CP violation seen in the kaon system actually arises from some
other mechanism.  One possibility which has been extensively studied
is spontaneous CP violation, when the vacuum and other physical states
are not CP eigenstates, even though the underlying Lagrangian may be
CP invariant.  In order for CP to be violated spontaneously the theory
must be more complicated than the minimal standard model, requiring at
least an additional Higgs doublet and the presence of certain
interactions between the Higgs doublets.  Of course, both CKM CP
violation as well as other sources of CP violation can be
simultaneously present in nature.  One of the main motivations for
building a ``B-factory'' -- an $e^+e^-$ collider optimized for
producing large numbers of B mesons -- is to determine whether CP
violation in the B meson system is that which is expected if a CKM
phase is responsible for the kaon CP violation\footnote{Measurement of
``direct'' CP violation in the kaon system is of great interest, e.g.,
observation of a non-zero value of the parameter $\epsilon'$ or certain
rare K decays, however strong interaction effects are theoretically more
difficult to handle than in the B system.}.  The non-observation of
neutron and electron electric dipole moments are powerful constraints
on non-CKM CP violation.

It is not clear whether CKM CP violation is large enough to account
for the observed bau, or whether there must be another source of CP
violation.  In order to address the question, one must consider some
particular mechanism for producing the baryonic asymmetry and do a
quantitative calculation.  Since this lecture is meant to introduce
the ideas to a general audience rather than be an exhaustive review, I
will focus on a baryogenesis mechanism which is applicable when the
boundary between high and low temperature phases (the bubble wall)
is thin compared to the scattering length of particles in the
plasma.  The mechanism in this case is simpler to understand
physically than for the thick wall case, and furthermore, recent
numerical work on the phase transition suggests that the wall is in
fact thin\cite{kajantie}.  In all mechanisms considered so far, the
fundamental crucial feature is that quarks and leptons couple to the
vacuum expectation value of the Higgs field in proportion to their
mass, since the spontaneous breaking of the electroweak gauge symmetry
is supposed to give rise to the masses of these particles.

Due to the interactions of the Higgs field with particles in the
high-temperature plasma, the effective potential of the Higgs field is
temperature-dependent.  As the temperature of the expanding universe
drops, the density of particles in the plasma decreases until finally
the effective potential is just the zero temperature potential of the
standard electroweak theory.  If there were no physics beyond the
minimal standard model, and if the Higgs mass were known, then all of
the parameters of the T=0 theory would be fixed.  The effective
potential at finite temperature would also be fixed, although its
computation is a highly non-trivial business when the Higgs mass is
not light, since non-perturbative effects become important\cite{s:condensate}.
It is known that at very high temperature the free energy of the
system is minimized when the vacuum expectation value (vev) of the
Higgs field vanishes\cite{kir,kirlin}.  On the other hand, in the T=0
theory the minimum energy occurs when the Higgs field has a vev of
$\sim 250$ GeV.  Thus the universe undergoes a phase transition at a
temperature which turns out to be of order $100$ GeV.

For sufficiently small Higgs mass, possibly consistent with the
present bound $m_H \gsi 60 $ GeV\cite{lep}, the phase transition will
be a first order transition, as is assumed in nearly all work on ew
baryogenesis. Then the picture is that as the universe cools and hits
the transition temperature, bubbles of the low-temperature phase
form.  These bubbles, in which the vev of the Higgs field is non-zero,
expand and eventually fill the universe.  The relevant CP violation is
supposed to occur as a result of the dynamics of quarks or leptons
interacting with the bubble wall -- the region in which the Higgs
field makes a transition from having zero vev to having a non-zero
vev.  Due to their thermal motion and the motion of the wall, quarks
and leptons present in the high temperature plasma near the wall will
encounter the bubble wall.  Since quarks and leptons get their mass
from their coupling to the Higgs vev, the more massive the particle,
the higher the potential barrier it sees to its motion.  For
instance, particles with insufficient kinetic energy cannot penetrate
into the low temperature phase and are totally reflected from the
bubble wall.  Thus CP violation in the interaction of a heavy particle
with the Higgs field will generally be more efficient for producing a
baryonic asymmetry than CP violation for a light particle.

The basic mechanism in the thin wall case was introduced by Cohen,
Kaplan and Nelson, who dubbed it the ``charge-transport'' mechanism.
It can be thought of in two steps.  First, particle scattering off the
phase boundary acts to {\it separate} some quantum number correlated
with left-chiral baryon or lepton number.  This gives rise to a
current of, lets say, left-chiral anti-baryon number toward the
unbroken phase (and correspondingly, since baryon number is not
violated in the interaction with the bubble wall, a left-chiral baryon
number current toward the broken phase).  Producing such a separation
requires CP violation in the interaction of quarks with the bubble
wall.  Some of the proposals for the origin of this CP violation are
discussed in the next section.  Given, say, a net flux of left-chiral
anti-baryon number into the unbroken phase, ew sphaleron transitions
in the unbroken phase can reduce the free energy by converting some of
the excess left-chiral anti-baryons into left-chiral leptons.  At this
point there is an excess of baryon number in the broken phase and
lepton number in the unbroken phase.  If the ew sphaleron transitions
are sufficiently suppressed in the broken phase, the excess of
left-chiral baryons present in the broken phase will {\it not} be
converted to anti-leptons.  A net baryonic density will remain after
the phase transition is complete, and is supposed to account for the
observed bau.  Note that since the ew sphaleron only acts on left
chiral particles and antiparticles, simply producing a separation of
left chiral baryon number is sufficient if it is large enough to
withstand erasure by the strong sphalerons as will be discussed in
section \S\ref{2higgs}.

In order that sphaleron transitions in the low temperature phase do
not equilibrate a baryonic excess created during the phase transition,
the sphaleron rate in the low-temperature phase must be {\it smaller}
than the expansion rate of the universe.  The rate of sphaleron
transitions in the broken-symmetry phase is approximately
\eq
\Gamma = T^4 \left(\frac{\alpha_W}{4\pi}\right)^4
N_{tr} N_{rot} \left(\frac{2E_{sph}(T)}{\pi T}\right)^7 \exp
\left(-\frac{E_{sph}(T)}{T}\right)
\label{sph_rate}
\en
where the factors $N_{tr} \simeq 26$ and $N_{rot} \simeq 5.3~10^3$ are
zero mode normalizations \cite{armc:zero} and
\eq
E_{sph}(T) \approx 3 M_W(T)/\alpha_W
\label{EsphMW}
\en
is the effective sphaleron mass at temperature $T$.  The sphaleron
rate will be less than the expansion rate of the universe if
\eq
E_{sph}(T_c)/T_c > 45.
\label{Esphlim}
\en
In the MSM, two parameters fix the Higgs potential.  The combination
of them which determines the $T=0$ vev is known, since it fixes $m_W$
at $T=0$.  If the mass of the Higgs were known, that would completely
fix the remaining freedom in the MSM Higgs potential.  The Higgs mass
is not known yet, but the requirement that the vev after the phase
transition is large enough to satisfy (\ref{Esphlim}) also constrains the
remaining parameter of the low temperature theory.  Taking the one-loop
effective potential for the Higgs field in the MSM, together with the
one loop approximation for the sphaleron rate, leads to the
requirement $\langle \phi(T_c) \rangle \gsi 2.4 g_w T$ and to the
bound $M_H < M_{crit} = 45$ GeV\cite{s:sm87}, which is inconsistent
with the present LEP limit of $\sim 60$ GeV.  Although the 1-loop
approximations to the effective potential have been improved by
resummation and calcuation of higher order effects, it is now clear
that the use of perturbation theory is inappropriate to extract this
information\cite{s:condensate}.  The reason is that in the high
temperature theory, the Higgs vev itself provides an essential
infra-red cutoff to perturbative calculations.  Thus perturbative
calculations are accurate in the broken phase, but break down in the
vicinity of the unbroken phase minimum, where the vev vanishes.
Recent lattice work indicates that the actual unbroken phase minimum
is much lower relative to the broken phase minimum than indicated by
perturbation theory, lowering the temperature of the phase transition
and modifying the bounds on the Higgs mass.  While it is numerically
difficult to find the new bound on the Higgs mass, it seems to be
consistent with the present experimental bound.  In the MSM it is
probably within reach of LEPII\cite{s:fkrs}, and even in a two-Higgs
doublet model it is likely that the lighter Higgs could be found at
LEP II.

In the next section we will discuss possible mechanisms for creating a
left-baryonic or leptonic current from quark or lepton interactions
with the bubble wall, which is the main topic of this talk.  However
before continuing with that, it is important to emphasize the great
quantitative uncertainty which surrounds the question of the
conversion of a given left baryonic current into a net baryonic excess
after the phase transition is complete.  In ref. \cite{fs:2} the
relation between the left baryonic current, $J_{CP}$, and the final
baryonic density was found in quasi-equillibrium approximation to be
$n_B = \frac{12}{5} J_{CP} f_{sph}$, where $f_{sph}$ is essentially
the probability that an excess antiquark will experience a sphaleron
transition before it diffuses into the broken phase.  $f_{sph} = 1$
when the dimensionless ratio $3 D_B \Gamma/ v^2$ is large
compared to one; $\Gamma$ is the sphaleron transition rate,
$D_B$ the baryonic diffusion constant, and $v$ the wall velocity.
However if this dimensionless ratio is much less than one, $f_{sph}$
is just proportional to it.  Unfortunately, estimation of $\Gamma$, the
sphaleron rate in the unbroken phase, has some 4-orders-of-magnitude
uncertainty due to its sensitivity to the non-perturbative physics
mentioned in the previous paragraph.  While ref. \cite{krs85} and
subsequent analytic and numerical work established that $\Gamma$ is
greater than the expansion rate of the universe at the time of the ew
phase transition, allowing sphaleron transitions to wipe out a
baryonic excess created in a GUT phase transition, it is not clear yet
whether $\Gamma >{ v^2 \over 3D_B}$.  Thus in all models there is a
substantial uncertainty in the predicted final asymmetry just due to
the uncertainty in the sphaleron transition rate.

\section{The basic source of CP violation}

In this section I will describe a number of possible sources for the
CP violation which produces the baryonic asymmetry, concentrating on
the ones which seem of greatest interest.  My goal is to elucidate the
fundamental issues, especially the consequences of CP, C, and P
invariance and gauge symmetries.  This leads me to examine the various
models from a somewhat different point of view than presented in the
original papers, so the reader interested in the perspective of the
authors of the ideas is urged to consult the original references.  A
number of new observations are made.  In order to make the amount of
material manageable, I concentrate exclusively on scenarios relevant
to thin bubble walls, as seem to b presently favored
theoretically\cite{s:condensate,s:fkrs}.  On account of the general
audience at the lecture, many of the details which follow were not
presented there.

It is useful to recall the transformation properties of vector and
axial vector fermionic currents under the discrete
symmetries\footnote{See, e.g., \S28 of ref. \cite{ll:rqt1}.}:
\begin{equation}
\begin{array}{lll}
{\rm C}: & J^{\mu} \rightarrow - J^{\mu} & J_5^{\mu} \rightarrow +
J_5^{\mu}  \\
{\rm P}: & (J^0,\vec{J}) \rightarrow (J^0,- \vec{J}) &  (J_5^0,\vec{J_5})
\rightarrow (- J_5^0,\vec{J_5})   \\
{\rm T}: & (J^0,\vec{J}) \rightarrow (J^0,- \vec{J}) &  (J_5^0,\vec{J_5})
\rightarrow (J_5^0,- \vec{J_5})
\label{cpt}
\end{array}
\end{equation}
Thus we immediately see that a baryonic density, $J^0$, violates C,
CP, and CPT, explaining why generation of the bau
requires\cite{sakharov} C and CP violation in the fundamental theory,
as well as a departure from thermal equilibrium, which can be thought
of as ``spontaneous'' CPT violation.  Note also that
$\vec{J_L}$ and $\vec{J_R}$ ($\equiv \vec{J} \mp \vec{J_5}$) are even
under CP, odd under CPT, and transform into the negatives of one
another under C.

As long as there are equal fluxes of quarks and antiquarks on
either side of the wall, CP violation in the reflection process is
needed to produce a left baryonic current, even though $\vec{J}_L$ is
even under CP.  Some of the left-chiral lepton or baryon number
separation mechanisms which have been considered are mentioned below.
They are generally based on the quantum mechanical reflection of quarks or
leptons from the interface between phases, such that the reflection
probability for a particle and its CP conjugate differ.  This arises
when there is an interference between the reflection phase shift (the
same for a particle and its CP conjugate) and a CP-violating phase in
the reflection process coming, for instance, from a non-trivial phase
in the coupling to the Higgs field.  In this case, the full reflection
amplitude for the particle is {\it not} just a phase rotation of the
amplitude for its CP conjugate and general it has a different
magnitude: $| a e^{i \phi} + b e^{i \delta}| \ne | a e^{i \phi} + b
e^{-i \delta}| $, where $\phi$ and $\delta$ are the CP conserving and
violating phases.

Even if CP is violated, CPT and unitarity imply that if the system is
in thermal equilibrium no net chiral baryonic or leptonic current is
established, because particles incident from opposite sides of the
wall make canceling contributions to the current and have the same
flux in equilibrium.  Of course during the phase transition the
bubbles of low-temperature phase are expanding, so that there is a net
flux of particles toward the unbroken phase and a baryonic or axial
baryonic current can be established.  We now turn to specific models
for producing the chiral current, starting with the simplest models to
analyze and progressing to the most difficult -- the minimal standard
model.

\subsection{``Singlet Majoron'' Model}
\label{majoron}

The earliest charge transport scheme \cite{ckn:L1,ckn:L2} involved
adding, a neutral heavy lepton, $N_R$, to the MSM.  CP is assumed to
be maximally violated in its Higgs coupling, in such a way that a
deficit of left-chiral lepton number develops in the unbroken phase
and a corresponding excess in the broken phase.  Then in the unbroken
phase, sphaleron transitions reduce the free energy by converting
left-chiral antileptons into left-chiral quarks.  The baryonic
asymmetry which arises in this mechanism is proportional to the
mass-squared of the heaviest of the light neutrinos, $\nu_{hl}$,
because the Higgs coupling of the lepton, which contains the CP
violation, is automatically proportional to its mass.  According to
the estimates of refs. \cite{ckn:L1,ckn:L2}, consistency with the
observed bau requires either the existence of a fourth generation or
requires $m(\nu_{hl})> 1$ MeV.  As a result of direct laboratory mass
limits on the three known neutrinos, only $\nu_{\tau}$ can be so
heavy.  However if the atmospheric neutrino oscillation ``hint'' holds
up and is explained by a $\nu_{\mu} \leftrightarrow
\nu_{\tau}$ oscillation, that would imply that $m(\nu_{\tau}) \sim
m(\nu_{\mu})\lsi 0.27$ MeV, ruling out this baryogenesis mechanism
without a fourth family.  Even without anticipating confirmation of
$\nu_{\mu}\leftrightarrow \nu_{\tau}$ oscillations, nucleosynthesis
constraints exclude a neutrino in the required mass range unless its
lifetime is $\lsi 100$s\cite{dolgovr}, so that this scenario seems
rather implausible now unless there is a fourth family.

\subsection{``Two Higgs Doublet'' Model and $Z$ condensation}
\label{2higgs}

\subsubsection{General Picture}

In the original two Higgs-doublet mechanism for baryogenesis, the top
quark is assumed to couple to a Higgs field whose vacuum expectation
value has a spatially varying phase inside the bubble
wall\cite{tz:prl,ckn:Y1,ckn:Y2}.  We can write its mass term as
\begin{equation}
\bar{t} m(z) e^{i \gamma_5 \theta(z)} t =
\bar{t} [m(z) cos \theta(z) + i \gamma_5 m(z) sin \theta(z)] t,
\end{equation}
where $\hat{z} \equiv \hat{3}$ is the inward pointing normal to the
bubble wall throughout this paper.  If $\theta(z)$ is a constant then
a global chiral transformation on the top quark field can be used to
remove it, so it has no physical effect\footnote{Note that if it were
not removed, the Dirac equation would look different for a particle
and its CP conjugate.  This serves as a reminder that one must be
careful to always compute physical quantities such as the net chiral
baryon current and not rely exclusively on apparent CP or parity
asymmetry of the Dirac equation.}.  However it is not sufficient for
$<\theta'> \equiv <\partial_z \theta(z)>$ to be non-zero, since this
is parity and CP invariant.  Hence physical consequences require a
non-vanishing $<\theta''>$.  We can ee this requirement directly as
follows: a chiral transformation can be used to remove $\theta$ in the
asymptotic broken phase where it is constant, and it can be set to
zero in the unbroken phase since with $m(z)=0$ there is no CP
violation independently of $\theta$.  Thus it must have a
non-vanishing second derivative to have any CP violating physical
effect.

Parity and CP violation in the interaction of the top quark with the
vev will in general result in different reflection coefficients for
left and right chiral top quarks, separating left chiral baryon number.
Baryon number itself is not separated because $\vec{J_L} + \vec{J_R}$ is C
odd while $\vec{J_L} - \vec{J_R}$ is C even, and this model violates
CP but not C, unless higher order corrections from gauge interactions
are also included.  Thus the net baryon number in either phase is
zero, with an excess of left chiral baryon number balanced by a
deficit of right chiral baryon number. The desired result is to have
the left-chiral baryon number, $n_B^L - \bar{n}_B^L < 0$ in the
unbroken phase and $>0$ in the broken phase, so that ew sphaleron
transitions in the unbroken phase convert some of the excess left
anti-baryon number to left lepton number.

We remarked above that thermal fluctuations in the SU(2) gauge fields
with non-trivial topology produce the electroweak sphaleron which
causes baryon and lepton number violating transitions.  In a similar
way, thermal fluctuations in the gluonic gauge fields with a
non-trivial change in the Chern-Simons number produce the ``strong
sphaleron'', violating quark chirality.  Since the strong sphaleron
has a faster rate than the ew sphaleron and it equilibrates left- and
right-chiral baryon number, it dilutes the expected final $n_B/s$
compared to the initial estimates\cite{ckn:Y2} which neglected this
effect. In the last year several papers have appeared on this
subject.  It was found that including strong sphaleron and other
effects reduces the final asymmetry by a large
factor\cite{giudice_shap,dine_thomas,jpt:constraints}. Fortunately for
this mechanism, this effect seems to be more or less compensated by a
more careful treatment of diffusion\cite{ckn:diffn} so that the latest
estimate\cite{ckn:diffn} still can be compatible with observation.

One can also use the two Higgs mechansim to produce a current of
$\tau_L$'s, which are not troubled by the strong sphaleron.  This
works best if the leptonic Yukawa couplings are large and the
relatively small $T=0$ lepton masses arise because $<v_2>$ is small at
$T=0$.  Ref. \cite{jpt:tau} concludes that this can be responsible for
the observed bau with a $\tau$ Yukawa coupling which is a factor $\sim
10$ larger than in the MSM.  Some investigation is required to see
whether such a large Yukawa, producing a vehicle for chirality change
even in the unbroken phase, could cause a dilution of the asymmetry as
the strong sphaleron does.

\subsubsection{Spatially Varying Phases and Gauge Invariance}

One has to be careful in treating gauge invariance in the two-Higgs
doublet model.  First let us review in greater detail how CP violation
in two-Higgs doublet baryogenesis works, and at the same time lay the
groundwork for a discussion of the MSM.  The relation to $Z^0$
condensation will emerge naturally.  Consider the part of the standard
model Lagrangian which involves quarks:
\eq
{\cal L} = {\cal L}_G + {\cal L}_Y.
\label{ewlagr}
\en
In the ``gauge'' basis,
\eq
{\cal L}_G = \bar{Q}_L {\not{\cal D}} Q_L + \bar{U}_R {\not{\cal D}}
U_R + \bar{D}_R {\not{\cal D}} D_R,
\en
and
\eq
{\cal L}_Y = \frac{g_W}{\sqrt{2}M_W} \{\bar{Q}_L^i V^{ij} M_d^{jj}
D_R^j\phi +
\bar{Q}_L^i M_u^{ii} U_R^i\tilde{\phi} + h.c.\},
\label{Yukawa}
\en
where ${\not{\cal D}}$ is the appropriate covariant derivative,
$Q_L^i$ are the left-handed quark doublets ($i$ is the generation
index), $U_R^i$ and $D_R^i$ are the right handed quarks with electric
charges $\frac{2}{3}$ and $-\frac{1}{3}$ respectively, $V$ is the
Cabibbo-Kobayashi-Maskawa (CKM) matrix, and $M_u$ and $M_d$ are the
diagonal mass matrices of the quarks.  In this basis, the Lagrangian has
been written in terms of the fields which are eigenstates of the
gauge interactions, and the CKM CP violation is contained in
a phase in the matrix $V$, relating the gauge eigenstates to the mass
eigenstates.  In the minimal standard model, $\tilde{\phi}_i$, the
Higgs which gives mass to the charge $2/3$ quarks $=
\epsilon_{ij}\phi^{\dagger}_j$. In a general 2-Higgs doublet model,
$\tilde{\phi}$ can be an independent doublet, or else the second
doublet can be taken to decouple from the quarks altogether and the
Yukawa couplings are as in the MSM.  In a the minimal supersymmetric
standard model (MSSM), supersymmetry requires that $\tilde{\phi}$ be a
distinct field from $\phi$.

The condition that the vacuum energy be minimized fixes the magnitudes
of the Higgs doublets and the relative phase between them.  In the MSM,
with a single Higgs doublet, one can always use the $SU(2)\times U(1)$ gauge
freedom to make the non-vanishing vev be real.  This is evident, since
any doublet can be written in the form
\eq
exp\left( i \frac{\sigma \cdot \xi(x,t)}{v} \right) \left(
\begin{array}{c} 0\\ \frac{v + \eta(x,t)}{\sqrt{2}}
\end{array}\right).
\label{vev}
\en
Taking $v$ to be the solution to the equation of motion for the
Higgs field, one then identifies $\eta(x,t)$ as the field
corresponding to the physical ``Higgs" particle.  At the critical
temperature, when the potential has degenerate minima, the vacuum
energy is minimized by a non-constant $v$, interpolating between the
vevs in the unbroken and broken phases.  We will work in the wall rest
frame\footnote{No important physics is affected if the vev in the
unbroken phase is taken to be extremely small but non-vanishing, so
that the singularity in (\ref{vev}) for $v \rightarrow 0$ is not a
source of problems for the arguments we wish to make.} where the vev is
static.

The most general $T=0$ potential normally considered in a two Higgs doublet
model can be put in the form
\begin{eqnarray}
V(\phi_1,\phi_2) = \lambda_1 (\phi_1^{\dag} \phi_1 - v_1^2)^2 +
\lambda_2 (\phi_2^{\dag} \phi_2 - v_2^2)^2 + \\
\lambda_3 [
(\phi_1^{\dag} \phi_1 - v_1^2) + (\phi_2^{\dag} \phi_2 - v_2^2)]^2
+ \lambda_4 [(\phi_1^{\dag} \phi_1) (\phi_2^{\dag} \phi_2) -
(\phi_1^{\dag} \phi_2) (\phi_2^{\dag} \phi_1)] + \\ \nonumber
\lambda_5 [ Re(
\phi_1^{\dag} \phi_2) - v1 v2 cos \zeta]^2 + \lambda_6 [ Im(
\phi_1^{\dag} \phi_2) - v1 v2 sin \zeta]^2, \nonumber
\label{potential}
\end{eqnarray}
with the $\lambda_i$'s real for hermiticity's sake.  This potential is
general enough to encompass the MSSM potential, and is only restricted
in a general non-susy model by having its dimension-4 terms invariant
under the discrete symmetry $\phi_1 \rightarrow -\phi_1$, a standard
method for avoiding large flavor changing neutral
currents\footnote{When $v_1 v_2 sin \zeta \ne 0$ this Lagrangian
explicitly violates CP.}.  Our first concern is to
insure that the vevs $v_1$ and $v_2$ only break $SU(2)\times U(1)
\rightarrow U(1)$ and not electromagnetism as well, since we used all
our gauge freedom in making the first vev be electrically neutral and
real.  The presence of the $\lambda_4$ term will insure this for a
large range of parameters, so we will henceforth assume that this has
been guaranteed.

The minimum of this potential can be taken to occur at
\eq
< \phi_1 > = \left(
\begin{array}{c} 0\\ v_1 \end{array}\right),
< \phi_2 > = \left(
\begin{array}{c} 0\\ v_2 e^{i \zeta} \end{array}\right).
\en
As long as the Lagrangian is {\it not} invariant under a global
redefinition of the relative phase between $\phi_1$ and $\phi_2$,
which would allow the $\zeta$ dependence to be removed, there is a
physically significant relative phase between the vevs in the $T=0$
theory.  This could be called ``vacuum CP-violation", reserving the term
``spontaneous CP-violation'' for a situation in which the Lagrangian
is CP invariant.  In a supersymmetric theory $\lambda_5 = \lambda_6$
in tree approximation, so that the two terms in the potential
containing $\zeta$ can be combined into a term proportional to
$\phi_1^{\dag} \phi_2 - v_1 v_2 e^{i \zeta}$ and there is no vacuum CP
violation.

At high temperatures the potential (\ref{potential}) receives
corrections from the interactions of the Higgs bosons with fermions
and gauge bosons and other Higgs particles in the plasma.  The main
effect is to introduce effective cubic self-interactions for $\phi_1$
and $\phi_2$, causing there to be two denererate minima in the
potentials for $|\phi_1|$ and $|\phi_2|$ without changing
qualitatively the $\lambda_5$ and $\lambda_6$ terms.  Thus inside a
bubble wall, where $|\phi_1|$ and $|\phi_2|$ are changing from their
unbroken to broken phase values, the equations of motion for $\phi_1$
and $\phi_2$ will in general produce a spatially varying relative
phase between them.

Now let us discuss the coupling of the Higgs fields to the quarks,
which is supposed to produce the quark masses and the CP violation in
the top quark or $\tau$ lepton scattering from the bubble wall.
Clearly, the $SU(2) \times U(1)$ gauge invariance which was used in
the discussion above to make $<\phi_1(x)>$ real can be used to remove the
phase from either $<\tilde{\phi}^0(x)>$, which by definition gives mass
to the charge $2/3$ quarks, or $<\phi^0(x)>$, which gives mass to the
leptons and the charge $-1/3$ quarks, even inside the bubble wall
where the relative phase between them is changing.  Yet this seems
paradoxical because we would have argued that if the CP violation is
in the top quark coupling, the bau thus generated would be $\sim
(m_t/m_b = 40)^2$ times greater than if it were in the bottom quark
coupling!  The resolution of this puzzle leads one naturally to the
subject of a $Z$ field condensate.

\subsubsection{Z Condensate}

Since we have excellent experimental evidence that Lorentz invariance
is unbroken, theorists generally never allow a field with a
non-trivial Lorentz behavior to have a vacuum expectation value.
However during the ew phase transition, the high-temperature plasma
provides a prefered Lorentz frame and the bubble walls break
translation invariance.  Thus one should consider the possibility that
a vector potential normal to the wall can have a vacuum expectation
value in the vicinity of the bubble wall.  For simplicity, it is
natural to assume that $SU(2) \times U(1) \rightarrow U(1)$, with
electromagnetism remaining a good symmetry even inside the bubble
wall.  Then only a vev for the $Z^0$ gauge potential need be
considered.  The gauge invariant quantities are $\zeta$, the relative
phase between the Higgs vevs $v_1 e^{i \theta_1}$ and $v_2 e^{i
\theta_2}$, and $Z_{\mu}^{GI} \equiv Z_{\mu} - \frac{2}{g}
(\frac{v_1^2 \partial_{\mu} \theta_1 + v_2^2 \partial_{\mu} \theta_2
}{ v_1^2 + v_2^2}) $.  Thus a complete specification of the vacuum
state requires not only specification of the magnitudes of the Higgs
vevs and the relative phase between them, but also specification of
the vev of $Z^{GI}_{\mu}$.  When {\it both} are specified, a $T_3$ or
hypercharge gauge change by an angle $\theta$ (say moving the phase of
the vev from the Higgs coupled to the top quark into the Higgs coupled
to the bottom quark) will not change the physical predictions of the
theory since it will induce a gauge condensate proportional to
$\partial_{\mu} \theta$, interaction with which produces the same
effect as the Higgs phase. Alternatively, in a particular gauge, one
must specify $Z_{\mu},~\theta_1$, and $\theta_2$ at all positions. The
original disucssions of the two Higgs doublet
models\cite{ckn:Y1,ckn:Y2} implicitly assumed $Z_{\mu} = 0$.

By the symmetry of the problem, $Z_{\mu}$ can only have a
non-vanishing component in the 3 or 0 direction.  These can only have
spatial derivatives in the $\hat{z}$ direction so that $\vec{B}^Z =
\vec{\nabla} \times \vec{Z} = 0$.  On the other hand, $\vec{E}^Z =
\vec{\nabla} \cdot Z_0 - \partial_0{\vec{Z}} $ can have a non-zero vev
as long as either $<Z_0>$ is non-zero and varying with $z$ or $
\partial_0{\vec{Z}} \ne 0$.  Turok and collaborators
\cite{jpt:classforce} recently have argued that the pure gauge
condensate $<Z_3> \ne 0$ can generate new physics.

The first step in figuring out the physical relevance of various
possible $Z$ condensates is to analyze their transformation properties
under CP, since a CP even quantity will not contribute to the
formation of a bau.  Since the Lagrangian is not parity invariant, the
gauge potentials cannot be characterized as vector or axial vector,
but when they have a constant vev they can be assigned a definite CP since
CP is conserved in the gauge interaction and the CP properties of the
currents to which they couple are determined.  From (\ref{cpt}) we see
that a constant $<Z_0>$ is CP odd while $<\vec{Z}>$ is CP even.  This
means that even if $<\vec{E}^Z> \ne 0$, it is CP even if it is a
constant and therefore has no leading order effect in producing a
baryonic asymmetry.  For the case of interest when the condensates are
spatially varying, we can consider their local Taylor series expansion
in $z$.  Then, e.g., $<Z_3'>$ is CP odd, etc.

Now let us turn to the proposal of a $<Z_3>$ condensate.  As discussed
above, gauge transforming a time-constant but spatially varying phase in
the Higgs vev produces a non-vanishing $<Z_3>$ which is
constant in time, so that the physics of such a condensate cannot be
any different than the physics of the gauge-equivalent 2-Higgs doublet
model with a vanishing $Z_3$ condensate.  Nonetheless it may be
advantageous to analyze the problem in a basis with $<Z_3> \ne 0$.
The most naive expectation in this basis would be that the CP
violating Bohm-Aharonov phase $\int Z_3 \cdot dz$ takes the place
of the CP violating Higgs phase.  However this may not be the case,
since in the $\theta$ basis a non-vanishing $\theta''$ is required in
order to have CP violation.  In the $<Z_3>$ basis this would
correspond to requiring $<Z_3'>\ne 0$.

Turok et al have recently argued\cite{jpt:classforce,jpt:thinwall} that in
the presence of a non-vanishing $<Z_3>$ there is a regime of momenta
in which fermions experience a CP violating ``classical force'' which
acts like a ``momentum filter'', and that this CP violation {\it does
not} depend on quantum interference between some CP violating phase
shift and a reflection phase shift.  As we shall discuss in section \S
\ref{qmreflection}, quantum mechanical interference effects may be
destroyed by collisions present in the high-temperature plasma, so a
mechanism which does not require interference is attractive.  However
there are still a number of features of this ``classical force''
proposal which need clarification.  First of all, it should be
emphasized that the starting point of using the Dirac equation
implicitly assumes coherence of the wavefunction, so that conclusions
following from such an analysis must be very carefully examined to make
sure that the presumed coherence plays no essential role.  Furthermore
refs.\cite{jpt:classforce,jpt:thinwall} explicitly drop terms in
$Z_3'$ in their WKB argument, while the general reasoning above would
suggest these terms are essential for actual CP violation\footnote{For
instance a $<Z_3>$ uniform in all space would just shift the zero of
the energy for left and right handed particles and shift their thermal
distribution functions, so the ``momentum filter'' effect would have
no physical significance.}.  Moreover, it is clear in the $\theta$
basis that the presence of the CP-conserving mass is essential to
having a genuine CP violation, while its relevance in the Turok et al
discussion is obscure.  In fact, in the usual analysis the spatially
varying CP conserving mass is necessary to allow quantum interference
with some CP-violating phase shift.  Thus understanding the dependence
on CP conserving mass in the $<Z_3>$ basis is necessary for
substantiating the claim that the a qualitatively new, ``classical
force'' has been discovered.  An actual solution to the Dirac equation
in the background of a specific $Z_3$ condensate which displays the
properties envisaged in refs. \cite{jpt:classforce,jpt:thinwall},
allowing computation of a physical current, could clarify these
issues.

Even if having a $Z_3$ condensate introduces nothing qualitatively
new with respect to coherence issues, retaining it explicitly may be
useful for determining the vevs in the vicinity of the bubble wall.
Nasser and Turok\cite{nt:msmcond} emphasize the importance of the
possibility of an instability in which a chiral top-quark pileup forms
in front of the wall.  The idea is that if some thermal fluctuation
produces a locally non-vanishing $<Z_3>$, top quark energy levels
would shift in such a way as to locally redistribute the $t_L$ and
$t_R$ densities, which in turn enhances the $<Z_3>$.  The mechanism
has only to do with the gauge couplings of the top quarks, so it is
independent of the Higgs sector and occurs in the minimal standard
model.  Since the ``seeding'' fluctuation is random, the sign of the
condensate and thus of its the contribution to the bau would vary from
region to region. In any given region of bubble surface the effect
could be large.  In order to get a non-zero result after averaging
over all regions, there must be some asymmetry in the amount of bubble
surface with positive and negative $<Z_3>$.  In the MSM, GIM
suppression (to be discussed in section \S\ref{msm}) would not appear
in the CP violation from a given bubble wall;  instead the local bau
production would resemble that of a two-Higgs doublet model.  The GIM
cancelation would manifest itself as a tendency for the $Z$ condensate
on different regions of bubble surfaces to produce bau's of opposite
signs to a very high degree of accuracy.  Thus the cruial issue
becomes the dynamics of competition between regions.

The idea of spontaneous CP violation having opposite signs on
different bubbles or different regions of the same bubble was
introduced by Comelli et al\cite{comelli} in the context of the
MSSM (see next section).  Using the usual simple description of
critical bubble formation in terms of the surface tension and energy
density between true and false vacuum, they estimated the final baryon
to entropy ratio to be the locally produced value of $\frac{n_B}{s}$
times $\frac{\Delta F}{T}$ where $\Delta F$ is the difference in free
energy of critical bubbles of the two types.  They obtained
$\frac{\Delta F}{T} = \frac{\Delta F(R_{\rm crit})}{T} \frac{3 \Delta
\sigma}{\sigma}$ where $\Delta \sigma$ is the difference in surface
tension for the two types of regions.  From ref. \cite{anderson_hall},
$\frac{\Delta F(R_{\rm crit})}{T} \sim 130$.  Nasser and
Turok\cite{nt:msmcond} give a qualitative discussion of the
competition between phases of $<Z_3>$ after bubbles collide, but it is
not explicit enough to allow comparison with the Comelli et al
estimate.

Work on this subject is only in its infancy and many important effects
have not yet been considered.  However if the general suggestion is
correct, and a dynamical instability in top quark reflection provides
an important contribution to the CP violating condensate on the bubble
wall, predictions of conventional models may be drastically altered.
The central difficulty in the analysis will become understanding the
dynamics of the evolution of the bubbles toward predominance of one
sign of spontaneous CP violation over the other.

\subsubsection{Minimal Supersymmetric Standard Model}
\label{mssm}

The supersymmetric minimal standard model is a special case of a
two Higgs doublet model, however as noted above, supersymmetry does
not allow the Higgs self-couplings necessary for the vevs to have a
non-trivial relative phase.  It was shown in ref. \cite{maekawa} that
at $T=0$ loop corrections involving soft susy breaking produce these
couplings and thus the possiblitiy of spontaneous CP violation.  Ref.
\cite{comelli} suggested that at high $T$ there can be spontaneous CP
violation such that the effective potential in the low-temperature
phase has two nearly degenerate minima, with phases of $\pm \pi$.
Thus on roughly half the bubbles $\theta_1(z)$ will decrease from 0 to
$-\pi$ in going from the unbroken to the broken phase, while in the
other half it will increase from 0 to $+\pi$.  On each bubble the
local bau production should be comparable to that of a maximal 2-Higgs
doublet model. Then they argue that a tiny explicit CP violation,
easily consistent with the limits on the neutron edm, could produce a
difference in surface tension on the two types of bubbles enough that
the net bau could be consistent with observation\cite{comelli}.

In the MSSM there can also be explicit CP-violating phases in the
couplings between gauginos (and higgsinos) and squarks, sleptons, and
Higgs and gauge bosons.  In the absence of some special circumstances,
these phases must be very small in order not to be in conflict with
the limits on the electric dipole moment of the neutron.  Production
of a bau via this explicit CP violation was investigated in the
``spontaneous baryogenesis'' scenario, appropriate when the bubble
wall is thick\cite{cn:mssm}.  It was concluded\cite{cn:mssm} that
if the perturbative effective potential is required to produce a
sufficiently strong phase transition (see section \S\ref{general}) and
the neutron edm is not too large, only a very tiny portion of parameter
space for the MSSM could give a large enough bau, and then only under
extremely optimistic assumptions.  However in view of the recent
developments regarding the importance of non-perturbative effects in
the effective potential\cite{s:condensate,s:fkrs} discussed in the
first section, the parameters of the MSSM should probably not be so
severely constrained as was done in \cite{cn:mssm}.  Furthermore, the
bubble wall seems likely to be better described as thin rather than
thick, so that explicit CP violation from the MSSM for the thin wall
regime needs to be studied.  In the this case, the explicit CP
violating phases in electroweak gaugino and higgsino couplings to the
Higgs vevs presumably produce a chiral higgsino and/or gaugino
current.  Since higgsinos and elctroweak gauginos have an $SU(2)_L$
anomaly, they couple to the sphaleron, so that a suitable asymmetry in
these could in principle lead to a baryonic excess.  This scenario
deserves quantitative investigation.  If it does not work,
baryogenesis in the MSSM will, like in the $Z$ condensate scenario for
the MSM, be dependent on understanding the difficult problem of
competition between phases and bubble evolution.

\subsection{The Minimal Standard Model}
\label{msm}

It is natural to ask whether the CKM CP violation which is usually
thought to account for the CP violation seen in the kaon experiments
can also account for the bau.  Since the MSM violates C as well as CP,
a baryonic current and not just an axial baryonic current can be
produced by the asymmetry in reflection probabilities, so that effects
of the strong sphaleron are less problematic than in two Higgs doublet
models.  However there is a simple argument which indicates that CKM
CP violation should have a practically negligible effect in cosmology.
To explain the argument we must first look more closely at CKM CP
violation.  In this model, CP violation is due to a non-trivial phase
in the matrix which relates the quark eigenstates for coupling to the
gauge fields to those for coupling to the Higgs field\footnote{Since
quark masses arise from their couplings to the Higgs field, the latter
eigenstates are just the physical mass eigenstates in the
zero-temperature theory.}. When there are just two generations this is
the familiar Cabibbo matrix, whose matrix elements can be taken to be
real and thus are CP conserving.  However using the freedom to perform
gauge and global chiral rotations on the quark fields, Kobayashi and
Maskawa showed that if there are three families of quarks, a general
CKM matrix is described by three Euler-like angles, plus a single
physically significant phase, $\delta_{CP}$, which cannot in general
be removed by rotations on the quark fields.  Not surprisingly, this
phase can be rotated away if any pair of the like-charge quarks are
degenerate in mass by using the extra freedom to rotate these
indistinguishable quarks into one another.  Similarly, if the gauge
and Higgs couplings of two generations are ``aligned'' (i.e., one or
more of the three CKM mixing angles vanishes) there is an additional
freedom to rotate them into one another, removing the CKM phase.

Now we can see why one might expect that CKM CP violation cannot be
responsible for generation of the bau.  CP violation vanishes when any
pair of quarks is degenerate or any CKM angle vanishes, in what is
known as ``GIM''(Glashow-Illipoulos-Maiani) cancellation.  Thus it
vanishes when
\begin{eqnarray}
\label{jarlskog_det}
& d_{CP} = sin(\theta_{12}) sin (\theta_{23}) sin(\theta_{13}) \sin
\delta_{CP}  & \\
& \cdot (m_t^2 - m_c^2)(m_t^2 - m_u^2)(m_c^2 - m_u^2)(m_b^2 -
m_s^2)(m_b^2 - m_d^2)(m_s^2 - m_d^2) & \nonumber
\end{eqnarray}
vanishes\footnote{The dependence on mass-squared differences rather
than just mass differences is due to the fact that the sign of a
fermion mass is not physically significant.}.  Since this is a
dimensional quantity, one might imagine that the temperature is the
natural dimension in the problem and estimate the dimensionless
``figure of merit'' for the effective magnitude of CKM CP violation
during the ew phase transition to be $d_{CP} T^{-12} \sim 10^{-18}$,
using experimental constraints on the product of sines of the CKM
angles and quark masses ($m_d \sim m_u \sim 0, ~ m_s \sim .15$ GeV,
$m_c \sim 1.6$ GeV, $m_b \sim 5$ GeV and $m_t \sim 175$ GeV) and $T
\sim 100$ GeV at the ew phase transition.

This estimate is not legitimate if the dependence on quark masses is
not perturbative in all the mass-squared differences, or if for some
reason the relevant dimensional parameter setting the scale is not the
temperature\footnote{Shaposhnikov\cite{s:msm} pointed out that it
also ignores the modifications to the effective local CKM matrix
arising from interactions of the quarks with gauge and Higgs particles
present in the high temperature plasma.}.  To assess the validity of
the $10^{-18}$ estimate, it is necessary to actually consider a
specific mechanism of baryogenesis.  However doing a correct
computation for this problem is much more difficult than in the models
discussed earlier, since in this case it is not enough to consider
simply the quantum mechanical reflection of a single species of quark
or lepton from the Higgs vev.  It is necessary to simultaneously treat
the reflection of all species of quarks, since with fewer than three
generations there is no CKM CP violation.  Furthermore the interaction
of quarks with the Higgs vev cannot by itself produce a baryonic
current in the MSM, since in the basis of mass eigenstates the CP
violating phase appears only in quark interactions with charged gauge
bosons.  Thus quark interactions with $W^{\pm}$'s in the plasma, as
well as with the vev of the neutral Higgs field, must be taken into
account.  However the interaction of a quark reflecting from the vev
is a non-local interaction, while the interactions of the quarks with
the gauge bosons in the plasma are local.  Until recently (see section
\ref{method} below) a formalism had not been developed for dealing
with this situation.

In order to get an idea of whether the bau produced in the MSM might
be significantly larger than $\sim 10^{-18}$, M. Shaposhnikov and I
took the following approach\cite{fs:1,fs:2}.  By working in the basis
of quasiparticle excitations of the plasma, some of the effects of the
interactions of the quarks with the gauge and Higgs particles in the
plasma are taken into account\footnote{The quasiparticles are
basically the quarks, ``dressed'' by their interactions with particles
in the high temperature plasma.}.  Then we considered the quantum
mechanical scattering of these quasiparticles from the Higgs vev.
This approach is not strictly consistent from the standpoint of
perturbation theory, since it includes for instance the $O(g^2)$
effect of interactions with $W$'s in the plasma on the quasiparticle
propagation, but not the effect at the same order of processes in
which real $W$'s in the plasma scatter from the quarks, while both
quarks and $W$'s may be reflecting from the wall.  On the other hand,
there is no obvious bias in neglecting the multi-body processes and no
method was known for including them, so that it seemed a reasonable
approach for getting an idea of the possible size of the CKM effect.

We found that in this approximation there is indeed a phenomenon for
which the perturbative estimate above is inapplicable.  In the thermal
plasma there is a spectrum of quark momenta normal to the bubble
wall.  For a quark of mass $m$, a fraction $\sim m/T$ of its phase
space corresponds to total reflection.  When the quark is strongly or
totally reflected, its interaction with the Higgs field is not at all
perturbative.  In particular, there is a region of momenta for which
the strange quark is totally reflected but the down quark is not,
in a fraction $\sim m_s/T$ of the total strange quark phase space.
In this region, the GIM cancellation which is at the heart of the
small result of the perturbative estimate is partially evaded, and one
finds a result which could be consistent with observation\footnote{One
can find the $s \leftrightarrow d$ and $s \leftrightarrow s$ quark
reflection amplitudes analytically in thin wall approximation,
perturbatively in the CKM mixing angles.  The quantity which replaces
$d_{CP} T^{-12}$ as the dimensionless measure of CP violation turns
out to be\cite{fs:2} $ \Delta(\omega) = -2 \left(\frac{\pi \alpha_W
T^2}{8 \omega M_W^2}\right)^3 \frac{m_t^4 m_c^2 s_{12}s_{23}s_{13}sin
\delta_{CP}}{ m_b^2 m_s} ~ Im(r_s)$, where $Im(r_s) \sim 1$ for $\omega$
such that the $s-$quark is totally reflected: $\omega \sim 50$ GeV.
This energy is much larger than the strange quark mass due to thermal
contributions to the quasiparticle mass gap.  See ref. \cite{fs:2} for
a discussion of the limitations of this expression for $\Delta$.}.
Solving the differential equations for the matrix of reflection
coefficients numerically, and attempting to include errors from all
sources, we estimated\cite{fs:2}
\begin{equation}
n_B/s \approx (10^{-9} - 10^{-12}) ~v ~f_{sph}~f_{3d}.
\label{result_vb=0}
\end{equation}
where $f_{3d}$ is the error introduced by having done a 1-dimensional
calculation rather than a three dimensional one, which we estimated to
be in the range $10^{-2} < f_{3d} < 10^{+2}$.  As noted above, the
sphaleron efficiency factor $f_{sph}$ enters the asymmetry calculation
in any model.  It is estimated\footnote{See, e.g., ref. \cite{fs:2}.}
to be in the range $10^{-4} - 1$, while the bubble wall velocity has
been estimated\cite{khl:wall,dlhll:pl,dlhll:pr,lmt:wall} to be $v \sim
0.1-0.9$.  Thus one can see that if circumstances are favorable, and
the quasi-particle-reflection approximation for computing the baryonic
current is a reasonable guide, that CKM CP violation could be
responsible for the observed bau, $\frac{n_B}{s} \sim (2-4) \times
10^{-11}$.

However it was stressed in ref. \cite{fs:2} that for the kinematic
region of importance in this mechanism, the quantum mechanical
reflection of quasiparticles does {\it not} provide a satisfactory
description of the problem since the penetration length of a
totally-reflecting strange quark into the broken phase is much larger
than the strong interaction collision length of the quasi-particle.
The implications of this will be discussed below in section
\S\ref{qmreflection}.  Another aspect of the calculation of
ref. \cite{fs:2} which should be examined critically is the
possiblitiy of what could be called a ``GIM conspiracy'', in which
there is a cancellation between processes in the purely electroweak
theory, for instance between processes involving real and virtual
$W$'s, and/or those involving real and virtual heavier quarks.  As an
example, $W$'s, Higgs, and top quarks outside the bubble are mostly
reflected back into the unbroken phase when they hit the bubble
wall. Thus in the plasma rest frame there is a net current of $W$'s,
Higgses, and tops toward the unbroken phase.  If the average momentum
transfer when left-chiral antiquarks scatter from one of these
particles, summed over boson charges and quark flavors, is different
than it is for left-chiral quarks, this would provide another
mechansim for generation of a baryonic current which could cancel or
add to the contribution from direct reflection.  Like the issue of
quasi-particle scattering which we discuss next, this is an aspect of
the baryogenesis problem which requires simultaneous treatment of
reflection and particulate interaction\footnote{When quark reflection
is not incorporated, the momentum-transfer asymmetry is negligible
because the perturbative argument applies.}.  A method for dealing
with such a situation will be presented in \S\ref{method}.

As a final remark on baryogenesis in the MSM, note that the phase
transition may be much more violent than indicated by the perturbative
calculations.  If bubble-expansion produces a region of ``compression"
in the surface of the bubble, with the Higgs vev taking on much larger
values than expected from the kink-solution to the perturbative
effective potential, even purely perturbative MSM CP violation
might account for the bau.  Since it depends (see eqn
(\ref{jarlskog_det}) ) on the twelfth power of the vev, a factor of 10
or so increase in the vev inside the wall could produce a bau of the
right order-of-magnitude.  This speculation is completely unmotivated,
but provides additional incentive for making a quantitatively accurate
theory of the phase transition.

\subsection{Breakdown of the Quantum Reflection Approximation}
\label{qmreflection}

In order for the validity of the quantum reflection approximation in
the high temperature plasma to be guaranteed, collisions with other
particles in the plasma must be irrelevant.  This is assured if the
collision length is large compared to the wall thickness, and also
compared to the penetration length of a totally reflecting particle.
For the quasiparticles of the high temperature plasma, the
mean-free-path for gluonic collisions is estimated to
be\footnote{\cite{fs:2} and references therein.} $\lambda_{inel} \sim
(0.15 g_s^2 T)^{-1} \sim 1/20~ {\rm GeV}^{-1}$, or up to a factor of
five larger for low-momentum quasi-particles when Debye screening is
taken into account.  Thus for wall thicknesses of order a few
$T^{-1}$, with $T\sim 100$ GeV, the former condition may be
approximately satisfied.  The latter condition is more problematic for
light quarks.  The calculation of reflection amplitudes in the
presence of total reflection relies on the boundary condition that the
rising exponential solution be discarded as unphysical since it is
non-normalizable.  This boundary condition is appropriate if the
particle which is totally reflecting has a sufficiently small
scattering probability off other particles in the medium that it is
unlikely to collide before the magnitude of the rising exponential is
signicantly greater than 1.  This is clearly satisfied if $ Im(p_z)
\lambda >> 1$, where $p_z$ is the $z$-component of the momentum in the
forbidden region and $\lambda$ is its mean-free-path.  However
although $|p_z| \sim 50$ GeV for the quasiparticle, the imaginary part
of its momentum in the broken phase is $\sim m(T)$, the product of the
Yukawa coupling of the quasiparticle and the vev in the broken phase.
For the strange quark $m_s(T) \sim 10^{-3} T$ so if the relevant mean
free path is $\lambda_{inel}$ for gluonic collisions given above, the
product $m_s(T) \lambda_{inel} \sim 1/200 - 1/40$.  Thus as stressed
in $\cite{fs:1,fs:2}$ for light quarks the quantum reflection
approximation cannot be used without additional justification.
On the other hand the approximation should be good for top quarks, for
which $m(T) \sim 175~ {\rm GeV} \frac{v_1(T)}{v_1(0)}$, unless
$\frac{v_1(T)}{v_1(0)}$ is very small\footnote{Following the
discussion in \S\ref{general}, suppression of the sphaleron rate in
the broken phase requires $ \sqrt{v_1(T)^2 + v_2(T)^2} \gsi 2.4 g_w
T$, but does not constrain $v_1(T)$ and $v_2(T)$ separately.}.  It
may also be good for the $\tau$ lepton, if the vev of the Higgs
coupled to the $\tau$ is large compared to its low temperature
value\cite{jpt:thinwall}, since the collision length is greater by a
factor $\sim (\alpha_{QCD}/\alpha_w)^2 \sim 10$.

When the collision length is not large compared to the penetration
length of the wave function into the forbidden region, it means that
the physical quantities of importance need to be computed directly,
without discussing reflection of single-particle states.  Indeed, the
whole notion of the reflection probability becomes meaningless because
there is no experimental way to prepare a particle far from the wall
and determine its probability of reflection.  Finding or not finding
an outgoing particle with the opposite momentum obviously is not the
relevant criterion in a plasma consisting of large numbers of
particles in random motion.  As an example, computation of the
``snowplow'' effect, in which quarks pile up in front of the bubble
wall impeding its expansion, is straightforward for sufficiently heavy
quarks because when $M^{-1} << \lambda$, quantum mechanical reflection is
the dominant mechanism since the collision length is small compared to
the penetration length.  On the other hand, the drag on the bubble in
a theory with 100 species of quarks having a mass $M/10 \lsi
{\lambda}^{-1}$ could be comparable, although in this case it would be
a complicated process with several mechanisms being important.  For
instance in addition to losing momentum as a result of their
interaction with the bubble wall, the quarks also collide with $W$'s
(which are efficiently reflected) and lose net momentum due to the $W$
flow away from the approaching bubble.

In the absence of a theory of how to treat these complicated
interactions, two groups\cite{ghop,hs} have tried to make models
to determine the result of a complete theory.  Both groups
retain the use of the single-particle QM reflection approximation,
although the details of their models differ.  In the quasi-particle
approach, collisions give rise to a ``lifetime'' or ``damping rate''
for the quasi-particle, corresponding to an imaginary part in the
quasiparticle propagator of order $\gamma \sim 0.15 g_s^2 T^{-1}$.
Gavela et al\cite{ghop} solve the Dirac equation with this imaginary
part to find the reflection coefficients of the ``decaying''
quasiparticles.  In order to avoid a manifest contradiction with
unitarity, they incorporate a ``source'' for the quasi-particles which
replaces them at a rate which keeps the system in equilibrium.  Using
this model, they find that the contribution of quasi-particle
reflection to the baryonic current $J_{CP}$ is many orders of
magnitude smaller than if there were no collisions.

While the existing calculations are clearly incapable of demonstrating
that the MSM accounts for the observed bau, the authors of refs.
\cite{ghop,hs} go further and claim that their model calculations
demonstrate conclusively that the MSM {\it cannot} be reponsible for
the bau.  I believe that this conclusion is too strong.  In a system
where multi-body collisions are crucial (as in the snowplow drag
computation mentioned above), the nature of the effect changes
sufficiently that one cannot draw any {\it a priori} reliable
conclusions from computing reflection coefficients, which is the
approach used in refs. \cite{ghop,hs}.  Moreover, both refs.
\cite{ghop,hs} {\it assume} that the effect of inelastic collisions is
to destroy the quantum coherence necessary to CP violation\footnote{In
ref. \cite{hs} this is an explicit assumption and in ref. \cite{ghop}
it is implicit in their mechanism of maintaining a constant
quasi-particle density even though quasi-particles decay.}. While they
regard this assumption as self-evident, I will argue below that it may
not be correct.  Since it is the crucial physical question it should
not be simply input as an assumption.

To show that one should not dismiss multi-body processes as
``obviously'' incoherent, let us consider in greater detail how the
baryonic current results from the interference between CP-violating
and CP-conserving phases.  The idea is to first look at some
particular quark reflection process (specifying, for instance, the
 momentum and angle of incidence and flavor of the quark) and then
examine the contribution to $J_{CP}$ of multigluon processes with a
similar quark current.  If all such processes contribute to $J_{CP}$
with the same sign, the multigluon processes would not tend to cancel
the contributions of the simple reflection, and the quantum mechanical
reflection approximation could give an estimate which is reliable up to
order $\alpha_s$ corrections.  Begin by considering a simple quark
reflection with some average baryonic current, $J_A$.  Let $A$ denote
the amplitude for this process, to zeroth order in $g_w$.  Due to the
interaction of the quarks with the bubble wall, $A$ will in general be
complex even at tree level.  The amplitude for the antiparticle
process will also be $A$ so the two contributions to $J_{CP}$ cancel.
Now consider the lowest order corrections to this amplitude which have
CKM CP violation, namely the orginal diagrams ``decorated'' with two
charged $W$'s.  Taking for the time being the CKM phase $\delta_{CP} =
0$, the full amplitude will be $A(1 + a e^{i \phi})$, where $a\sim
g_w^4$ is defined to be a positive real number.  The phase $\phi$ is
non-zero because the masses of the quarks in the intermediate state
are different than in the leading term and therefore lead to a
different reflection phase shift.  When the CKM phase $\delta_{CP}$ is
reinstated, the amplitude will be $A(1 + a e^{i (\phi +
\delta_{CP})})$; the amplitude for the same process in which the
quarks are replaced with antiquarks of the same chirality is $A(1 + a
e^{i (\phi - \delta_{CP})})$.  The net contribution to the baryonic
current from the quark and antiquark processes will be proportional to
$J_A a sin \phi ~sin \delta_{CP}$.  Now consider a similar process,
but with some number $n$ of external gluons, which makes a contribution
$J_B$ to the baryonic current with the same sign as the original
$J_A$.  Parametrically, $J_B \sim g_s^{2 n} J_A$.  When the virtual
two-$W$ corrections to this process are included, we can write its
amplitude as $B(1 + b e^{ (\phi' \pm \delta_{CP})})$ for the particle
and antiparticle process respectively.  When we add this to the
contribution from the first process we find a total baryonic current
$(J_A a sin \phi + J_B b sin \phi') sin \delta_{CP}$.  Since the
virtual short distance $W$ exchange is the same in the two cases and
$\phi$ and $\phi'$ originate from interaction with the bubble wall, we
expect $b sin \phi' \sim a sin \phi$ and thus both processes
contribute with the same sign to the net baryonic current, even though
the amplitudes $A$ and $B$ are obviously incoherent.

We have seen that as long as the $W$ exchanges in the two processes
are short distance effects, the multi-gluon processes will contribute
coherently to $J_{CP}$ even though there is obviously no phase
relation between the amplitudes involving different numbers of gluons.
However when the $W$'s in the loops go on shell there are additional
phases from the loop integrals and one can no longer argue that $b sin
\phi' \sim a sin \phi$.  This suggests that the flavor decoherence
length of the strange quark, $\lambda_{fd} \sim \frac{\alpha_s
M_W^2}{alpha_W m_c^2} \lambda_{inel}$\cite{fs:2}, may be the relevant
length scale determining when the quasiparticle reflection
approximation can be used for estimating the baryonic current.

The discussion given above is very crude and could not convince anyone
that gluonic effects do not wash out a CKM contribution to the bau.
Clearly one needs a systematic approach to the problem, which can
consistently keep track of the relative signs of the contributions of
different processes to the net baryonic current in order to tell what
the net effect of the multitude of different processes will be.
However this example shows that one should not make the a priori {\it
assumption} of refs. \cite{ghop,hs} that the multi-body processes
contribute with random signs to the net baryonic current relative to
one another.  In order to discard with confidence the MSM as the
source of the bau, one must make a first principles calculation of the
effect of multi-body processes.  An outline of a method to do this is
given in the next section.

\section{Field Theory in the Background of the Bubble Wall}
\label{method}

In this section, I will describe a framework for systematically
studying aspects of the physics which involve both local particulate
scattering and non-local quantum mechanical reflection.  The material
is technical and was therefore only covered in summary form in the
lecture at the ``Trends in Astroparticle Physics" workshop.  It is
essential for reliably treating problems in which the collision length
is comparable to or smaller than $m^{-1}$, as is the case in the MSM
model in the important region of strange quark total reflection.

\subsection{General Discussion}

The ultimate quantity of interest for determining the bau produced
during the electroweak phase transition is the expectation value of
the baryonic charge in the broken phase, long after the wall has
passed.  In the plasma rest frame it is the 0-th component of the
four-vector
\eq
<J_{\mu}(x)> = \sum_i <\bar{\Psi}^i(x) \gamma_{\mu} \Psi^i(x)>,
\label{current}
\en
where the superscript labels each type of quark.  Thermal and
non-equilibrium effects are included by taking the appropriate
ensemble average, innocuously denoted $<...>$.  Up to now this density
has been computed for the charge-transport mechansim by the procedure
described in the previous sections:
\begin{enumerate}
\item  Find the reflection coefficients for quarks and antiquarks of
both chiralities from the bubble wall.
\item  Find the current $J_{CP}$, by determining the contribution of
each species and chirality of quark and antiquark to the current of
interest, assuming fluxes from each side of the wall corresponding to
equilibrium statistical distributions coming from the two phases,
boosted to the wall rest frame.
\item  Consider the action of the sphaleron on the system with the
current $J_{CP}$ flowing into the unbroken phase to determine the
final baryonic density.
\end{enumerate}
A conceptually better approach is to instead develop field theory in
the background of the non-constant vev.  In perturbative
approximation, the current (\ref{current}) gets a non-vanishing
contribution already from a simple quark loop in the two-Higgs doublet
baryogenesis model, but is only non-vanishing at 3-loops with CKM CP
violation, as will be shown below.  The effect of quark and gauge and
Higgs boson scattering from the bubble wall is included {\it
nonperturbatively} by using propagators in the background of the
changing vev.  A given higher order diagram for (\ref{current})
corresponds to a number of physical processes which can occur in the
background of the wall.  For instance the diagram with a quark loop
and two exchanged $W$'s represents quarks interacting with virtual
$W$'s while propagating to or from the wall, as well as the multibody
reflection process $Q^i + W \rightarrow Q^j + W$.  Thus this approach
naturally overcomes the difficulty with the quasi-particle reflection
approach, of not being a consistent expansion in the coupling by not
including multi-body scattering-with-reflection\footnote{This is not a
conceptual problem for the two-Higgs doublet model because in that
case CP violation is present simply from the interaction with the vev,
so all higher order corrections can be ignored without losing the
effect.}.  Equally importantly, it allows the effect of higher-order
gluonic interactions to be reliably assessed because local particulate
scattering as well as non-local reflection from the wall can both be
included.

In order to obtain a non-zero value for $<J_0>$, sphaleron
interactions must be included.  In principle, they can be treated as a
(space dependent) insertion in the fermion line.  The dependence
of the sphaleron rate on the local Higgs vev means that the insertion
is small or negligible in the broken phase.  Its value on a given
quark line will depend on the chemical potentials of the other
left-chiral quarks and leptons in the local environment.  The steady
state of the entire system should in principle be determined
self-consistently, fixing the wall velocity and individual particle
distribution functions for every flavor of quark and antiquark, lepton
and antilepton, gauge and Higgs particle, as a function of position
relative to the wall, in terms of each other.  However a much simpler
approach, which employs the approximation already in use for
charge-transport calculations, is to ignore sphaleron insertions in the
calculation of $<J_{\mu}(x)>$ and deviations of the particle
distribution functions from what they would be if the vev were
constant with its local value. Then given $J_{CP} \equiv
<\vec{J}(x)>$, one can take into account the sphaleron processes and
diffusion through the bubble wall by using the results developed in
section 5 of ref.\cite{fs:2} relating $n_B$ to $\frac{12}{5} J_{CP}
f_{sph}$.

To implement this approach, one must first find the Greens functions
for free quarks and gauge and Higgs particles in the background of the
changing vev.  The quark propagators in the background of the changing
Higgs field depend in a non-trivial way on the quark Yukawa couplings
even in the unbroken phase, since the solutions to the Dirac equation
in the presence of the vev contain the reflection amplitudes for
scattering from the Higgs field.  For instance for a step function
vev, the reflection amplitude for a scalar is $\frac{k-l}{k+l}$ where
$k$ and $l$ are the momenta in the unbroken and broken phases.  The
propagators also contain a non-trivial CP-conserving phase because the
reflection coefficient is complex for at least some incident
momenta\footnote{In the thin wall limit it is complex for the
regions of integration in which total reflection occurs: $k^2 <
m^2$, while for finite wall thicknes it can be complex even without
total reflection.}.  The zero-temperature momentum-space quark
propagator in a theta function background vev is given in ref.
\cite{glop}.  In the presence of the wall, the scattering solutions in
the gauge and Higgs boson sector are changing superpositions of the
asymptotic fields and one must retain a finite wall thickness if one
wishes all the asymptotic states to have finite mass.  Analytic
solutions are given in ref. \cite{fm:1}; this allows the free field
operators of the standard model bosons to be written down and their
propagators to be determined.  Since energy and momentum parallel to
the bubble wall is conserved, but translation invariance perpendicular
to the wall is lost, calculations are simplest in a mixed
representation in which propagators depend on $z,~z',~\kperp$ and
$\omega$, where $z$ and $z'$ are the initial and final coordinates
normal to the bubble wall.   Equilibrium finite temperature
perturbation theory works just as usual and finding the temperature
Green's functions is straightforward.  The program is nearly complete
to this point for the standard model and will be reported
elsewhere\cite{fm:inprep}.  Note that it is entirely different from
the approach of Gavela et al\cite{ghop} which considers only
quasi-particle reflection and thus does not reap the benefits of the
field theoretic formalism.

As noted previously, $J_{CP}$ vanishes when fluxes from the broken and
unbroken phases are equal as they are in thermal equilibrium.  Thus it
is not sufficient to use equilibrium finite temperature field theory
to solve this problem.  However the approximation conventionally used
for charge-transport baryogenesis can be easily implemented.
Particles incident from the broken phase are taken to have
distribution functions appropriate to particles in equilibrium with
the plasma deep in the broken phase, and correspondingly for those from the
unbroken phase, all boosted to the wall rest frame.  Although it may not
be possible to put the corresponding ``temperature'' Green's functions
into a compact form, that is not necessary to performing the calculations.

Given this framework, the importance of gluonic interactions can be
assessed by computing the gluonic corrections to the leading
contribution.  Although this requires a 4-loop calculation for the
MSM, the general question can be addressed in the 2-Higgs doublet
model.  Given an analytic solution to the Dirac equation with a spatially
varying phase for the vev, one can find the Greens function for the
quark in the 2-Higgs doublet model.  Computing the
$m_q \rightarrow 0$ part of the 1-gluon correction to the basic quark
loop should be a feasible calculation.  An indication of trouble from
gluonic interactions would be the presence of collinear or soft
logarithms.  If these are detected, the techniques developed
for studying Sudakov suppression in QCD can be extended to ascertain
whether resummation of the higher order QCD corrections damps the
leading order effect.  In the absence of such logs or some other
indication of a breakdown in perturbation theory when the height of
the barrier is small, the strong corrections will be nothing but an
order $\alpha_s$ correction to the leading electroweak result.

An important virtue of the approach advocated here is that it
automatically includes all effects, systematically at any given order
of perturbation theory, with the correct relative weights and
coherence properties.  It solves the problem of accounting for both
coherent quantum mechanical reflection and the localized interacions
of particles which are themselves being reflected and transmitted from
the interface between low and high temperature phases.

\subsection{Application to the MSM}

Consider computing the expectation value of the current, using
perturbation theory to whatever order is necessary.  In order to
produce a net non-vanishing result for $J_{CP}$, an interference is
needed between $\delta_{CP}$ and the CP conserving phases appearing in
propagators and in loop integrals (which can have a non-vanishing
absorptive part due to the presence of real intermediate states).  In a
2-Higgs doublet model, the CP violating phase would already be present
in the quark propagator, as well as a CP-conserving phase associated
with the reflection process, so that a non-zero result for $J_{CP}$
should appear already in the basic quark loop.  On the other hand, in
the minimal standard model, a non-zero result for $J_{CP}$ arises
first at 3-loop order.  This can be seen as follows.  Taking $T=0$
quark mass eigenstates as our basis, quark interactions with the
neutral Higgs are purely diagonal in flavor and thus the Green's
functions are flavor-diagonal.  Unlike the two-Higgs doublet model
with a spatially varying vev, the phases in the MSM propagators are
exclusively CP-conserving.  The couplings of the quarks to $W^{\pm}$'s
and charged Higgs bosons are proportional to the CKM matrix $V$, and
in this basis at least 2 $W^{\pm}$'s or charged Higgs bosons are
required in the loop in order fo there to be a non-trivial dependence
on $\delta_{CP}$.  With a single $W$ exchanged in the quark loop, the
flavor structure is just $tr[V A V^{\dagger} B]$. The antiparticle
contribution is given by replacing $V \rightarrow V^*$. But $tr[V^* A
V^{t} B] = tr[B V A V^{\dagger}] $ because $A$ and $B$ are diagonal in
the basis we have chosen.  This is equal to the particle contribution
by the cyclic property of the trace, so there is no CP violation until
next order.

Non-trivial flavor dependence, necessary if the GIM cancellation is to
be evaded in the MSM, arises from the mass dependence of the quark
Green's functions and from Higgs vertices.  In the quasi-particle
reflection approximation, it was shown in ref. \cite{fs:2} that the
cancellation is evaded in the region of momenta in which the $s$ but
not $d$ quark is totally reflected.  Now we can ask whether there
could be a ``GIM conspiracy'', with cancellations occuring between
different processes in the purely electroweak theory.  There are four
distinct diagrams with two charged $W$'s exchanged in the quark loop,
as well as diagrams with charged Higgs bosons replacing the $W^{\pm}$'s.  In
each of these diagrams the structure in flavor space is $tr[V A
V^{\dagger} B V C V^{\dagger} D]$, where $A,~B,~C,~D$ are diagonal but
not proportional to the unit matrix\footnote{Since $V V^{\dagger} =
1$, those portions of $A,~B,~C,~D$ which are proportional to the unit
matrix give vanishing contribution to $J_{CP}$.}.  In the absence of
the wall, the only way for $A,~B,~C,~D$ to have a non-trivial flavor
dependence is for there to be Higgs interactions on the quark lines
between the $W^{\pm}$ vertices.  But since Higgs interactions change
the chirality of the quark, and the $W^{\pm}$'s couple only to left
chiral quarks, one would need two Higgs interactions to change the
chirality from $L$ to $R$ and back to $L$, so that $A,~B,~C,~D \sim
M^2$ and one arrives at the perturbative estimate.  In fact, if the
only source of flavor dependence is from the Yukawa couplings at Higgs
vertices, one can see that there is no CP violation whatever at 3-loop
order, as follows.  In this case, since the propagators in the loops
are being taken to be massless, for every contribution to $A$ there is
an identical contribution to $C$ taking, e.g., $A$ and $C$ to be
associated with the $D-$quark lines.  But $tr[V A V^{\dagger} B V C
V^{\dagger} D]+ tr[V C V^{\dagger} B V A V^{\dagger} D]$ is the same
as the antiparticle contribution obtained by replacing $V \rightarrow
V^*$, taking the transpose and using the cyclic property of the trace.
This is the essential content of the argument of ref. \cite{s:m^14}.

In the actual problem, however, modifications in the quark propagators
due to the presence of the wall introduce the possibility of a less
suppressed dependence on some of the mass matrices between $V$'s and
$V^{\dagger}$'s.  For instance in a non-normal collision with the
wall, angular momentum conservation does not require chirality flip as
it does for normal incidence.  At the same time, the reflection
coefficients are still dependent on the height of the wall, and thus
are flavor dependent.  While it is possible to identify contributions
which can have a sufficiently favorable dependence on quark masses to
possibly account for the bau, a detailed investigation is necessary to
estimate the final result.  The complete calculation necessary to
obtain a quantitatively accurate result and see whether the predicted
sign is correct, is in principle straightforward but extremely
difficult.  Fortunately it should be possible to answer the most
urgent question, whether gluonic collisions obliterate the production
of a baryonic or axial baryonic current, with a much easier
calculation in the two-Higgs model.

\subsection{Corrections to the electroweak sphaleron}

Near the bubble wall, there can in principle be non-trivial CKM-dependent
corrections to the fermionic lines in the effective vertex which lead
to a different rate between sphalerons involving quarks and those
involving antiquarks.  Naively taking the 't Hooft effective
interaction in the basis in which gauge interactions are diagonal, and
adding three Higgs to form a closed loop and gluons to fix the color,
one can find diagrams which can interfere with the lowest order
process. In the absence of the vev and the non-trivial mass dependence
thus introduced into the quark propagation, it is easy to show that
there is no CP violation arising from these corrections.  However when
the mass dependence is more subtle, this may not continue to be the
case.  Moreover the excess of top quarks in front of the wall, and the
overall antisymmetry in flavor and color of the quarks in the sphaleron
effective interaction, means that the sum over ways to insert the three
Higgs into the flavor structure of the sphaleron is not a symmetric sum
as it is in the usual trace over flavors.  It does not seem
particularly likely that including these corrections will give a
significant contribution to the bau, however the possiblity should be
looked at more carefuly before being discarded.

\section{Summary}

After reviewing in the first section the general issues of
baryogenesis, the second section was devoted to discussing
possible sources of the CP violation needed to produce the observed
baryonic excess at the electroweak phase transition.  The emphasis was
on making the physical ideas of some of the most interesting
possibilities clear, not on giving a complete review of all models.
Special attention is given to the relation between two-Higgs doublet
models and the question of $Z$-condensation, and to the problems of
baryogenesis in the MSM.  The supersymmetric minimal standard model is
seen to be surprisingly similar to the non-supersymmetric minimal
standard model in spite of having a more complicated Higgs sector and
other possible sources of CP violation.  It was argued that in order
to decide whether the MSM can be responsible for the bau, it is
necessary to compute and include the contributions of multi-body
processes, since pure quantum mechanical reflection is negligible for
light particles in the high temperature plasma.

The final section outlines a new approach to treating dynamics near
the bubble wall when both the non-local interaction of particles
with the changing vev, and their interaction with other particles in
the plasma, may be important.  It provides a framework for settling
the question of the importance of gluonic corrections to the
electroweak processes in the MSM, and allows the issue of a possible
GIM conspiracy to be investigated.  The methodology presented here of
doing perturbation theory in the background of the changing vev can
also be used to obtain improved particle distribution functions in the
neighborhood of the bubble wall\cite{fm:inprep}.

No ``bottom line'' on electroweak baryogenesis is given.  While a
number of models have promise of explaining the observed baryonic
asymmetry, it could also be the case that no model works.  Many
details regarding the phase transition, sphaleron rates, and bubble
dynamics need to be understood better before any firm conclusion will
be possible.

\section{Acknowledgements}
I have benefitted from discussions with many colleagues, especially G.
Baym, M. Losada, A. Masiero, J. McIntosh, V. Rubakov, M. Shaposhnikov,
S. Somalwar and S. Thomas.


\end{document}